\documentclass{PoS}

\usepackage{graphicx}

\def\Tr{\mathop{\rm Tr}}

\def\LL{\left\langle}   
\def\RR{\right\rangle}  

\def\BE{\begin{displaymath}}
\def\EE{\end{displaymath}}
\def\BEA{\begin{eqnarray*}}
\def\EEA{\end{eqnarray*}}
\def\BNEA{\begin{eqnarray}}
\def\ENEA{\end{eqnarray}}

\newcommand{\Chi} {\raisebox{0.4ex}{$\chi$}}
\def\LL{\left\langle}   
\def\RR{\right\rangle}  

\title{The 2+1 flavor topological susceptibility from the asqtad action at 0.06 fm}

\ShortTitle{2+1 flavor topological susceptibility at 0.06 fm}

	\author{C.~Bernard\\ 
	Physics Department, Washington University,
        St. Louis, MO 63130, USA}
	\author{B. Billeter, C. DeTar, and L. Levkova\\
        Physics Department, University of Utah, Salt Lake
        City, UT 84112, USA}
	\author{Steven~Gottlieb\\
        Physics Department, Indiana University,
        Bloomington, IN 47405, USA}
	\author{U.~M.~Heller\\
        American Physical Society, One Research Road, Box
        9000, Ridge, NY 11961-9000, USA}
	\author{\speaker{J.~E.~Hetrick}\\
        Physics Department, University of the Pacific,
        Stockton, CA 95211, USA\\
	E-mail: \email{jhetrick@pacific.edu}}
	\author{J.~Osborn\\
	Boston University, 
        Boston, MA 02215, USA}
	\author{D.~B.~Renner and D.~Toussaint\\
        Physics Department, University of Arizona, Tucson,
        AZ 85721, USA}
	\author{R.~Sugar\\
        Physics Department, University of California, Santa
        Barbara, CA 93106, USA}

\abstract{
We report new data for the topological susceptibility computed on 2+1
flavor dynamical configurations with lattice spacing 0.06 fm,
generated with the asqtad action. The topological susceptibility is
computed by HYP smearing and compared with rooted staggered chiral
perturbation theory as the pion mass goes to zero. At 0.06 fm, the raw
data is already quite close to the continuum extrapolated values
obtained from coarser lattices. These results provide a further test
of the asqtad action with rooted staggered flavors.}

\FullConference{The XXV International Symposium on Lattice Field Theory\\
		 July 30-4 August 2007\\
		 Regensburg, Germany}

\begin{document}

\section{Introduction}

In 2001 S. D\"urr presented \cite{Durr} an analysis of the
dependence of the topological susceptibility on the pion mass, as
measured in then current full QCD simulations. 
His comparison included results using Wilson fermions from CP-PACS,
UKQCD, and SESAM/T$\chi$L, as well as results using thin link
staggered fermions from the Pisa group, and by A. Hasenfratz
who analyzed MILC and Columbia dynamical lattices. The
conclusion from these studies was that simulations were not yet in
agreement with chiral perturbation theory \cite{chipert1} which says
that (using $f_\pi$ = 130 MeV)
$$
\Chi_{\rm topo} \sim \frac{f_\pi^2 m_\pi^2}{4N_f}
$$ as the pion mass tends to zero. While in most simulations there was
a reduction in $\Chi_{\rm topo}$ as $m_\pi^2$ is reduced, contact with
the above line was largely absent. This is displayed here in figure 1,
reproduced from \cite{Durr}, with the above expectation (linear in
$m_\pi^2$) shown as the left black line, against the data. Also shown
is the quenched $m\rightarrow \infty$ expectation as a horizontal line on
the right.
\begin{figure}[h]
\begin{center}
\includegraphics[width=12cm]{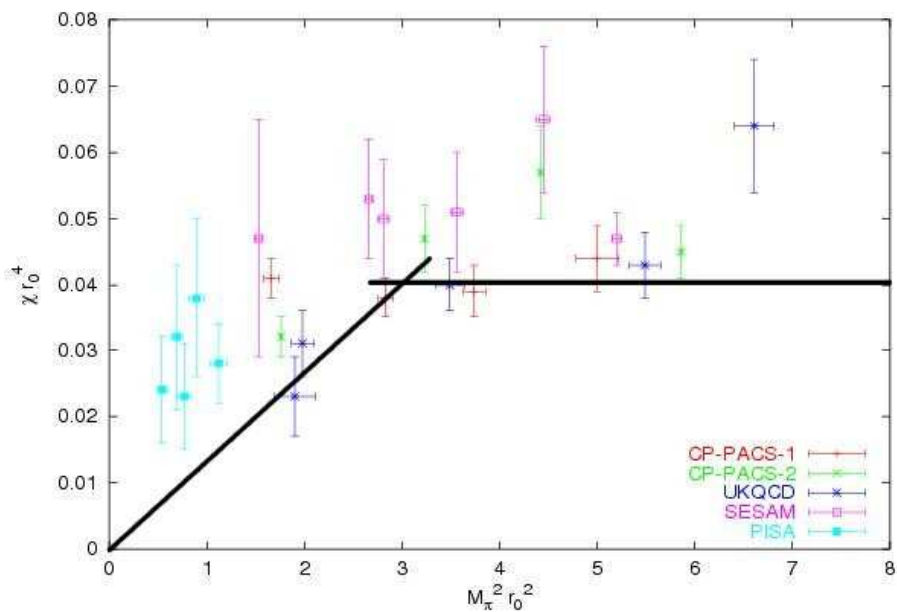}
\end{center}
\caption{ Comparison of full QCD data for $\chi_{\rm topo}$ ca. 2001,
taken from \cite{Durr}.}
\end{figure}

About the same time, improved actions were seeing a renaissance and
have since generated striking results in almost all areas of lattice
gauge theory.

In this contribution we present the latest results of the MILC
collaboration for the topological susceptibility and its mass
dependence as $m\rightarrow 0$, using the Asqtad action
\cite{asqtad} at a lattice spacing of 0.06 fm. With 
this improved action, HYP smearing, a variance reduction technique
to determine the topological charge, and careful extrapolation to the
continuum limit using rooted staggered chiral perturbation theory, we
indeed see encouraging agreement with the expectations from QCD.

\section{Simulations}

By 2003 the MILC Collaboration had lattices at two lattice spacings,
$a = 0.12$ and $a = 0.09$ fm, and a variety of masses with which to 
investigate the dependence of the topological susceptibility on quark
(or equivalently pion) mass; these results were presented in
\cite{firstTopo} and are shown below in figure 2.
\begin{figure}[h]
\begin{center}
\includegraphics[width=8cm]{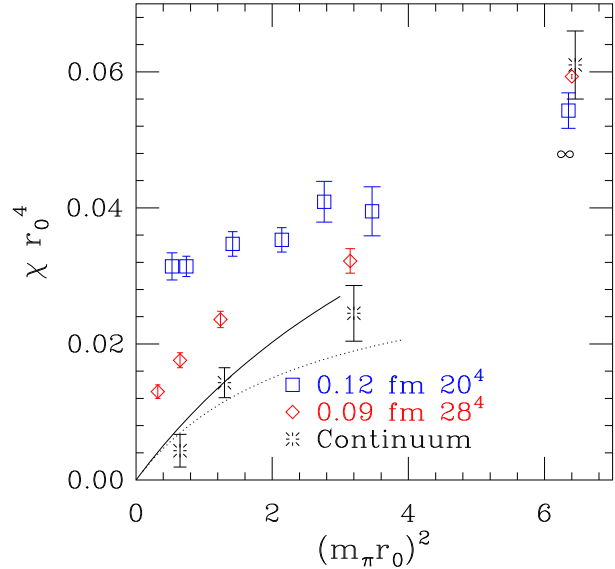}
\end{center}
\caption{MILC data from \cite{firstTopo} (2003).}
\end{figure}

This result was quite encouraging, as the expectations of chiral
perturbation theory appear to be supported by the data.

In the last 4 years, much larger lattices (up to 48$^3 \times$ 144) at a
lattice spacing of 0.06 fm have been generated and their topological
charge analyzed. It is the purpose of this note to update the above
picture with the new results. In addition to now having three lattice
spacings, there have also been some theoretical developments in the
methods used to extrapolate continuum values which we will report as
well.

\subsection{Topological charge and susceptibility measurements} 

As in \cite{firstTopo}, we continue to measure the topological charge density
$F_{\mu\nu}\tilde F^{\mu\nu}$, using the Boulder extended link
definition and three HYP smearing sweeps \cite{HYP}. 

Typically the topological susceptibility, $\Chi_{\rm topo} = \LL Q^2
\RR/V$, is computed by averaging the individual $Q^2$ from each lattice
in the ensemble ($V$ is the lattice volume). For our study, we have used a
method introduced in \cite{LAT04} which significantly reduces the
variance in $Q^2$. Since 
$$
\Chi_{\rm topo} = \frac{\LL Q^2 \RR}{V} = \int dr \LL q(r) q(0) \RR
$$
we measure the correlator $\LL q(r) q(0) \RR$, which we split into a
short distance ($r < r_{\rm cut}$) part and a long distance part
($r < r_{\rm cut})$. $r_{\rm cut}$ is $\sim$8-10 in lattice units.

At short distance, where the correlator is large, we use the measured
points in computing $\LL q(r) q(0) \RR$, whereas at long distance
since the measured correlator has large variance, we use values
obtained by a fit to points in the transition region (using the green
points in the example shown in figure 3). The fit is to a Euclidean
scalar propagator, $m K_1(mr)/4\pi^2 r$, which is the expected long
distance behavior of $\LL q(r) q(0) \RR$.
\begin{figure}[h]
\begin{center}
\includegraphics[width=13cm]{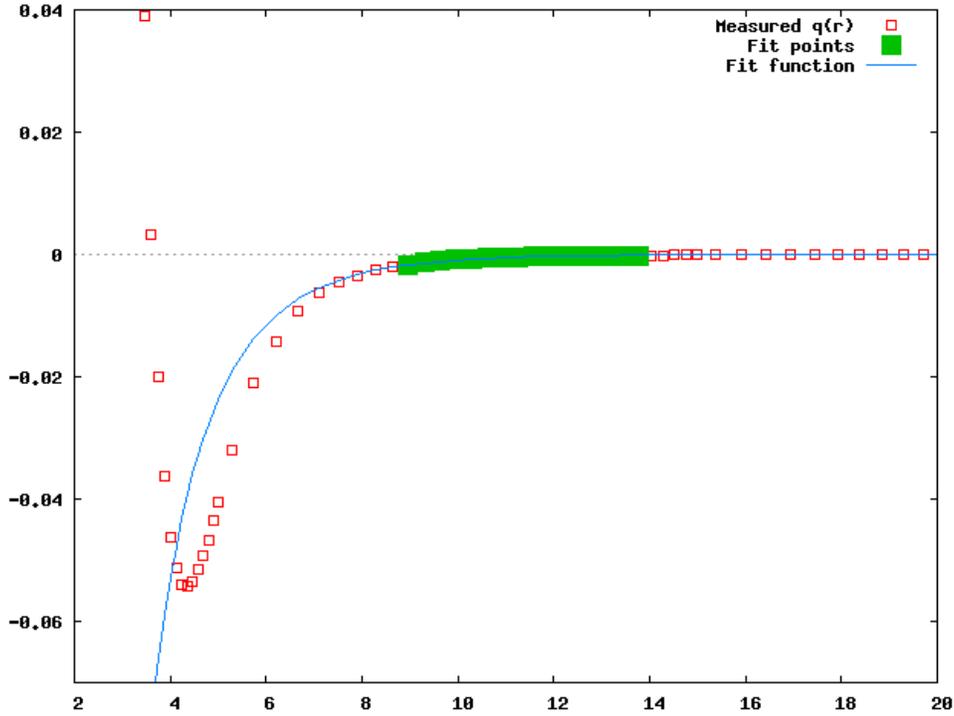}
\end{center}
\caption{Points used to compute $\LL q(r) q(0) \RR$. Measured points
at $r < r_{\rm cut} \sim 9a$ are used. For $r > r_{\rm cut}$ values
from a fit to those points in green are used.}
\end{figure}

\subsection{Staggered chiral perturbation theory}

In addition to finer lattices and a variance reduction method for our
$\Chi_{\rm topo}$ measurement, there have also been improvements to
the expectations from chiral perturbation theory. In particular there
is now good analytic understanding of the taste splittings induced in
operators at finite lattice spacing. \cite{LeeSharpe, AubinBernard}.
An example of these taste splittings in the staggered pion multiplet
is shown in figure 4. The masses of the 16 pions of various tastes are
shown as the quark mass is taken to zero. At finite lattice spacing the
splitting is seen to be roughly constant. As $a\rightarrow 0$ these
pions become degenerate.
\begin{figure}[h]
\begin{center}
\includegraphics[width=8cm]{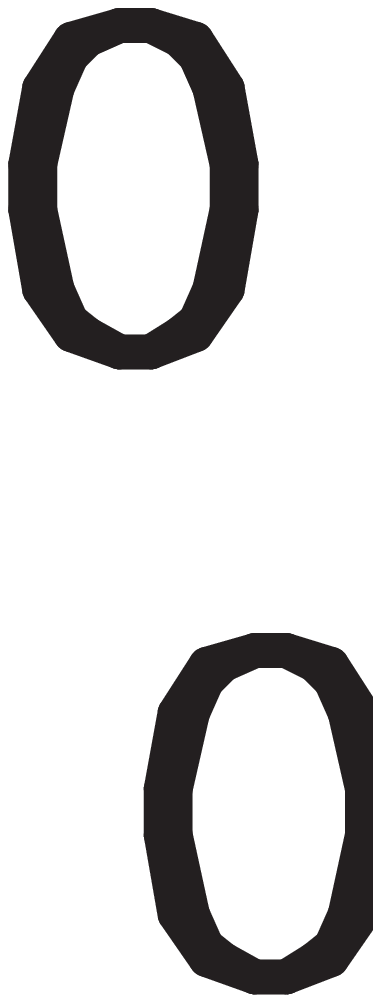}
\end{center}
\caption{Taste splittings of the staggered pion taste multiplet as a
function of quark mass} 
\end{figure}
The residual $U(1)$ chiral symmetry of staggered fermions provides one
Goldstone pion in the multiplet (shown in the figure above with black
diamonds), whose mass goes to zero with vanishing quark
mass, even at nonzero lattice spacing. 
Typically it has been this pion which is studied in lattice
QCD, as was the case in our previous topological susceptibility
study \cite{firstTopo}.

In \cite{BDO} however, it was recognized that the staggered pion field
which couples to the anomaly, and hence should be used for the
relevant chiral perturbation theory, is the the {\em taste singlet}
pion, with mass $m_{\pi,I}$, as opposed to the commonly used Goldstone
pion (a pseudoscalar in taste). This taste singlet pion is the one
displayed with magenta diamonds in figure 4, at the top of the
multiplet.

Following \cite{LeeSharpe, AubinBernard} and starting from the 
rooted staggered chiral lagrangian
$$
{\cal L} = \frac{f_\pi^2}{8}\Tr(\partial_\mu U^\dagger\partial U)
  - \frac{\mu f_\pi^2}{4}\Tr[{\cal M}(U^\dagger + U)]
+ \frac{m_0^2}{2}\phi^2_{0I} + \sum C_i {\cal O}_i  +\dots
$$
where $\frac{m_0^2}{2}\phi^2_{0I}$ is an explicit mass term
representing the coupling of the
anomaly to the taste singlet pion field $\phi_{0I}$,
Billeter, Detar, and Osborn derive \cite{BDO} the following dependence of 
$\Chi_{\rm topo}$ on pion masses in the 2+1 flavor case:
\begin{equation}
\Chi_{\rm topo} = \frac{f_\pi^2 m_{\pi,I}^2/8}{1 
+ m^2_{\pi,I}/(2m_{\bar{s}s,I}) 
+ 3m^2_{\pi,I}/(2m^2_0)}
\end{equation}
where $m_{\bar{s}s,I}$ is the $\bar{s}s$ taste singlet pseudoscalar
meson mass.
This formula interpolates smoothly between the $m_{\pi,I}^2
\rightarrow 0$ chiral limit:
$$
\lim_{m\rightarrow 0}\Chi_{\rm topo} \sim \frac{f_\pi^2 m^2_{\pi,I}}{8}
$$
and the quenched limit:
$$
\lim_{m\rightarrow \infty}\Chi_{\rm topo} 
\sim \frac{f_\pi^2 m_0^2}{12} \approx 0.06/r_0^4
$$
In the last formula, we use the measured quenched topological
susceptibility to set the value of $m_0$, and we repeat that 
$m_{\pi,I}$ here is the mass of the taste-singlet pion.

\section{Results}

Putting these developments together, we present our latest results for
the 2+1 flavor topological susceptibility. The lattices used for this
study have taken more than five years to produce and analyze, 
and are shown in Table 1.
\begin{table}[t]
\begin{center}
\begin{tabular}{|l|l|c|l|r|}
\hline
$am_{u,d}$ / $am_s$  & \hspace{-1.0mm}$10/g^2$  & lattice spacing 
& $L^3\times T$ & \# lats. \\
\hline                             
quenched       & 8.00  & a = 0.12 fm & 20$^3\times$64 & 408   \\
0.05  / 0.05   & 6.85  & & & 425    \\
0.04  / 0.05   & 6.83  & & & 351    \\
0.03  / 0.05   & 6.81  & & & 564    \\
0.02  / 0.05   & 6.79  & & & 484    \\
0.01  / 0.05   & 6.76  & & & 658    \\   
0.007  / 0.05  & 6.76  & & & 493    \\   
\hline
quenched        & 8.40  & a = 0.09 fm & 28$^3\times$96 & 396  \\
0.031  / 0.031  & 7.18  & &  & 496  \\   
0.0124  / 0.031 & 7.11  & &  & 527  \\   
0.0062  / 0.031 & 7.09  & &  & 592  \\   
\hline
0.0072 / 0.018  & 7.48  & a = 0.06 fm & 48$^3\times$144 & 624 \\
0.0036 / 0.018  & 7.47  & &  & 608 \\
\hline
\end{tabular}
\end{center}
\caption{Lattices and parameters used in this study} 
\end{table}
Having $\Chi_{\rm topo}$ at numerous quark masses and three
lattice spacings, we fit our entire data set to an interpolating
function in lattice spacing, $a$, and taste singlet pion mass squared, 
$m_{\pi,I}^2$ (the strange taste-singlet mass $m_{\bar{s}s,I}$ on
these lattices was tuned to be constant):
$$
\frac{1}{\Chi_{\rm topo}r_0^4}(m^2_{\pi,I}, a)
            = A_0 + (A_1 + A_2 a^2 + A_3 a^4)/m^2_{\pi,I}.
$$ 
The continuum limit is obtained from the fit by setting $a=0$ in
this function, and we are left with 
$\Chi^{\rm cont.}_{\rm topo}(m^2_{\pi,I})$ extracted from our data. The result
is shown below in figure 5.
\begin{figure}[h]
\begin{center}
\includegraphics[width=10cm]{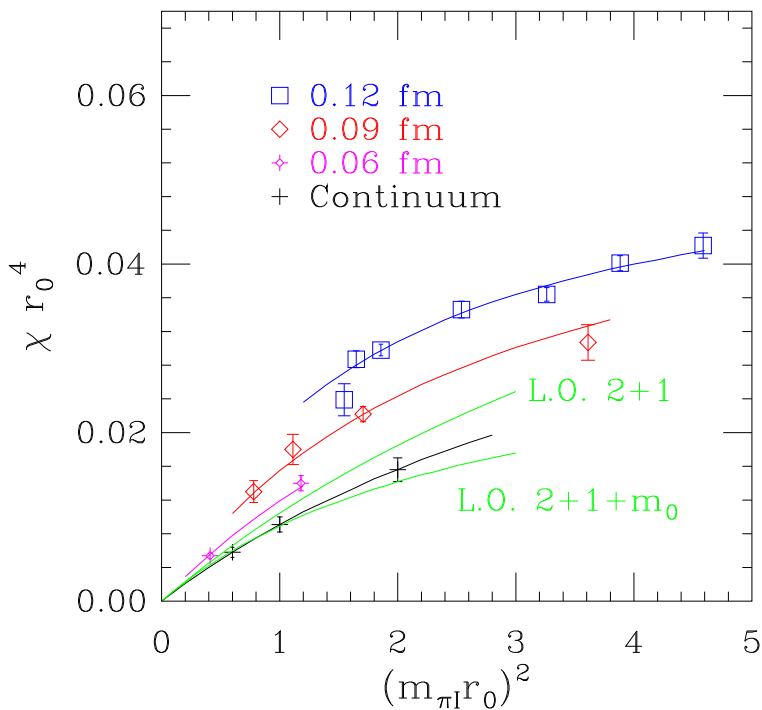}
\end{center}
\caption{Latest MILC results for the topological susceptibility from
lattices down to $a=0.06$ fm.} 
\end{figure}
Measured lattice data are shown with blue, red, and magenta symbols,
while the continuum limit extrapolation function $\Chi^{\rm cont.}_{\rm
topo}(m^2_{\pi,I})$ is shown with a solid black line. Some
representative points along this line are shown with error bars
reflecting the errors of the continuum extrapolation. Finally,
two functions representing the chiral perturbation prediction of eq. (2.1)
are shown in green: the lower line ``L.O. 2+1+$m_0$'' includes the value for
$m_0$ set by the quenched data, and eq. (2.1)
with $m_0 = \infty$ is shown labeled ``L.O. 2+1'' for comparison. 

\section{Conclusions}

With the addition of the new $a=0.06$ fm data, we see that the
topological susceptibility is behaving as expected in the
$m^2_{\pi,I}\rightarrow 0$ limit of rooted staggered chiral perturbation
theory. We find it striking that the lightest 0.06 fm datum is almost
on the continuum line without extrapolation.

Finally, we feel that these results lend further credibility to the
use of the ``fourth root method'' to simulate single
flavors. As mentioned in M. Creutz's talk at this conference, 
aberrant results from the fourth root would be expected to arise first in
violations of topological quantities and correlations, which are quite
sensitive to the number of flavors. We see no such violations, and
indeed only strong support that the simulations are behaving as
expected from QCD.

\end{document}